\documentclass{aa}
\usepackage{graphics}

\newcommand{\am}{$'$}
\newcommand{\as}{$"$}

\begin{document}

%%%%%%%%%%%%%%%%%%%%%%%%%%%%%%%%%%%%%%%%%%%%%%%%%%%%%%%%%%%%%%%%%%%%% TITLE
\thesaurus{08.01.2;
           10.15.2 IC\,4651;
           13.25.5}
\title{Soft X-ray emission from intermediate-age open clusters: IC\,4651}
%\subtitle{      }
\author{ T. Belloni\inst{1} \and G. Tagliaferri\inst{2} }
\offprints{T.~Belloni  }

\institute{   Astronomical Institute ``Anton Pannekoek'', 
              University of Amsterdam and Center for High-Energy Astrophysics, 
              Kruislaan 403, 1098 SJ Amsterdam, The Netherlands
         \and 
              Osservatorio Astronomico di Brera, Via Bianchi 46,
              I-22055 Merate, Italy
                }
\date{Received ??, accepted ??}

\titlerunning{Soft X-ray emission from intermediate-age open clusters: IC\,4651}
\authorrunning{T. Belloni \& G. Tagliaferri}

\maketitle

%%%%%%%%%%%%%%%%%%%%%%%%%%%%%%%%%%%%%%%%%%%%%%%%%%%%%%%%%%%%%%%%%% ABSTRACT

\begin{abstract}
We present the results of soft X-ray observations of the intermediate-age
open cluster IC\,4651 performed with the ROSAT PSPC. We detected 25 sources.
Two are identified with a giant binary and a blue straggler respectively, both
belonging to the cluster, and two with probable main-sequence members.
Two other cases have ambiguous identification.
Of the five binaries known in the cluster, the one detected in X rays
is the only one whose period is short enough to maintain fast rotation
and therefore strong stellar activity at a high age.
The detected blue straggler is probably binary, suggesting
that binarity is the key to producing a high level of X-ray emission.
It is the third blue straggler detected in X rays. 
The remaining sources need to be identified through optical follow-up.
\end{abstract}

\section{Introduction}

Stellar activity depends crucially on the star's rotation rate
(Pallavicini et al. 1981), which decreases with age because
of magnetic braking (e.g. Skumanich 1972). Thus, open clusters are crucial
to distinguish between truly evolutionary effects on stellar activity
and effects primarily due to the rotation rate itself. The first X-ray
observations of open clusters were carried out with the {\it Einstein}
satellite (Stern et al. 1981, Caillault \& Helfand 1985,
Micela et al. 1988, Schmitt et al. 1990,
Micela et al. 1990), but these observations have been carried out
in a more systematic way with the ROSAT satellite
(Stern et al. 1992; Stauffer et al. 1994; Patten \& 
Simon 1993; Randich \& Schmitt 1995; Randich et al. 1995, 1996a,b).
Various clusters of different ages have been studied in order to understand
the evolution of stellar activity with age (see Randich 1997 for a review).
Before ROSAT, the attention
has been concentrated mainly on young clusters (30 to 700 Myr). 
ROSAT performed the first observations of the old open cluster M\,67
($\sim$5\,Gyr, Belloni et al. 1993, Belloni et al. in preparation) 
and NGC\,188 ($\sim$9\,Gyr, Belloni et al., in preparation), 
leading to the detection of a number of sources. Indication
of chromospheric activity has been found for most of the optical candidates
(Pasquini \& Belloni 1994).  In contrast to younger clusters, clusters
older than $\sim$\,1 Gyr are not expected to contain rapidly-rotating 
single late-type stars, and therefore strong X-ray sources. 
However, there are stars older than $\sim$ 1 Gyr that show rapid rotation:
these are members of close binary systems, where tidal interaction prevents
the stars from losing angular momentum; well-known examples are the RS CVn
binaries. The observations of these clusters have led to the detection
of such binaries, and also to a number of
peculiar and interesting objects, such as white dwarfs, catalclysmic variables,
blue stragglers and wide/eccentric binaries.

In the framework of a project to cover the X-ray observational gap between
old and young clusters, we observed the intermediate
age open clusters NGC\,752 ($\sim$2\,Gyr, Belloni \& Verbunt 1996), NGC\,6940
($\sim$1\,Gyr, Belloni \& Tagliaferri 1997) and IC\,4651 ($\sim$2.5\,Gyr)
with ROSAT PSPC.
Moreover, ROSAT HRI observations of the clusters NGC 3680 ($\sim$2\,Gyr)
and NGC\,2527 ($\sim 1$\,Gyr) were made in 1997 and are currently being
analyzed.
In the PSPC observation of NGC\,752, 49 X-ray sources have been detected;
seven of them are identified with optical cluster members, four of
which are short period binaries, one is a rapid rotator and one is
a blue straggler (Belloni \& Verbunt 1996). 
In the PSPC observation of NGC\,6940 18 sources were detected,
four of which are identified with members of the cluster with a 
fifth source a suspected member. In NGC\,6940, a high fraction
of the detected members are binaries: three out of four of the identified
members are among the only six binaries known in the cluster. These 
observations give also evidence for the presence of
a saturation level, at which the whole surface of the star is 
chromospherically active (see Belloni 1998 for a review).

Here we present the results obtained for IC\,4651. This open cluster has
an estimated age of $\sim 2.5$\,Gyr and a distance of $\sim 800 - 900$\,pc
(Eggen 1971, Anthony-Twarog et al. 1988). It has a relatively constant
reddening, estimated to be E(B-V)$ = 0.15$ by Eggen (1971) and E(B-V)$ = 0.09$
by Anthony-Twarog \& Twarog (1987), and an angular size smaller than 
$\sim 15'$. The paper is organized as follows: in Sect. 2 we present the PSPC 
observation and our data analysis, in Sect. 3 we present and discuss
the results and in Sect. 4 we compare them with those of other open
clusters and discuss the implications.

\section{Observations}

We observed IC\,4651 with the Position Sensitive Proportional Counter
(PSPC) on board ROSAT between 1993 Sep 9th 06:05 UT and 1993 Sep 10th
18:30 UT, for a net observation time of 15200 s. A description of the
satellite and the instrument can be found in Tr\"umper (1983) and
Pfeffermann et al. (1986) respectively.
The data were analyzed using the EXSAS package (Zimmermann et al. 1994). 
Due to the degradation of the Point Spread Function at
large off-axis angles, we limited our analysis to the inner 20$'$ of the
field of view. We followed the standard procedure within EXSAS for the
detection of sources. First we produced a background map by removing all
possible sources and smoothing the resulting image. Then we ran
a Maximum Likelihood (ML) technique to test for deviations from
a purely background distribution (Cruddace et al. 1988).
The ML threshold for detection was set at 10, corresponding to a 
single-trial probability of a chance detection of $4.5\times 10^{-5}$.
We applied the procedure described above for three
PSPC channel bands: 11--240 (total band T, corresponding roughly to 0.1--2.4
keV), 11--40 (soft band S, 0.1--0.4 keV) and 41-240 (hard band H, 0.4--2.4 keV).

With this procedure, we detected 24 sources in the T band, 26 in the 
H band and 3 in the soft band. A few of these sources turned out to
be either double detections (in some case the ML program detects the same
source more than once) or very broad excesses due to inhomogeneities
of the background (which are too broad to be point sources and too
weak to be statistically significant extended sources).
After exclusion of these sources we
cross-correlated the three lists, producing a final list of 25
sources detected in the inner 20$'$ of the PSPC detector.
A summary of the sources is given in Table 1. The reported positions have
been corrected for the offset ($\sim 10''$) between the X-ray detector and the
optical star sensor by means of a cross-correlation with stars in the
Space Telescope Guide Star Catalog (GSC: Lasker et al. 1990). 
Since eight sources could be identified with GSC entries, the boresight 
correction obtained with this procedure is robust and allows us to 
remove all systematic effects. Therefore, we added only a systematic
error of 3$''$ to the 90\% error radii, in order to account for
residual uncertainties.

\begin{table*}
\caption[]{Sources detected in the IC\,4651 field. The columns give 
source ID number, position, 90\% confidence radius, 
count rate, channel band to which the count rate 
corresponds (T=11--240, H=41--240), hardness ratio HR2 (see text)
for the brightest sources (above 100 detected counts),ID of the optical object
(L: Lindoff 1972; E: Eggen 1971; AT: Anthony-Twarog et al. 1988; OUT: outside
the optical fields studied by the authors mentioned above),
$V$ and $B-V$, X-ray luminosity in the 0.1--2.4 keV band estimated
with a typical X-ray spectrum of RS CVn (see text), and remarks.}
\begin{flushleft}
\begin{tabular}{rccccccccccl}
     no.                     &
     $\alpha$(2000)          &
     $\delta$(2000)          &
     $\Delta r$              &
     Rate                    &
     B                       &
     HR2                     &
     Opt.                    &
     $V$                     &
     $B-V$                   &
     L$_X$                   &
     remarks                 \\

                             &
                             &
                             &
                             &
(cts/ks)                     &
                             &
                             &
ID.                          &
                             &
                             &
(erg/s)                      &
                             \\
\hline
 1  & 17h23m15.0s &-49$^\circ$56\am 17\as &  8\as & 8.4$\pm$ 1.5 & T &
--0.27$\pm$0.16&OUT  & 10.60   &       &         & GSC      \\
 2  & 17h23m38.2s &-50$^\circ$ 0\am 16\as & 11\as & 3.0$\pm$ 0.6 & T &
               &OUT  &  9.36   &       &         & SAO 227893 \\
 3  & 17h24m11.4s &-49$^\circ$44\am 26\as & 17\as & 1.8$\pm$ 0.5 & H &
               &OUT  & 10.74   &       &         & GSC      \\
 4  & 17h26m08.0s &-49$^\circ$44\am 53\as & 18\as & 5.4$\pm$ 0.8 & H &
               &OUT  &$\sim$16 &       &         & POSS     \\
 5  & 17h23m48.3s &-49$^\circ$45\am 50\as & 12\as & 3.3$\pm$ 0.6 & H &
               &OUT  & 11.72   &       &         & GSC      \\
 6  & 17h24m22.8s &-49$^\circ$46\am 40\as & 12\as & 2.6$\pm$ 0.5 & H &\\
 7  & 17h25m 4.0s &-49$^\circ$47\am 36\as & 15\as & 1.4$\pm$ 0.4 & H &
               &OUT  &$\sim$14  &       &         & POSS     \\
 8  & 17h24m56.5s &-49$^\circ$50\am  4\as &  4\as &11.3$\pm$ 0.9 & H &
0.36$\pm$0.08 &L241 & 10.97   & 1.18  & $1.8 \cdot 10^{31}$ & P=75d, e=0.09 \\
 9  & 17h23m36.7s &-49$^\circ$53\am 34\as & 17\as & 1.3$\pm$ 0.4 & H &\\
10  & 17h25m29.5s &-49$^\circ$54\am  7\as &  8\as & 2.7$\pm$ 0.5 & H &
              &L234 & 14.14   & 0.63  & $4.3 \cdot 10^{30}$ &               \\
11  & 17h24m44.8s &-49$^\circ$54\am 58\as & 13\as & 1.7$\pm$ 0.4 & H &
              &L11  & 11.5    & 0.7   & $2.7 \cdot 10^{30}$ &               \\
"   &             &                       &       &              &   &
              &L12  & 11.9    & 0.4   &        "             &               \\
"   &             &                       &       &              &   &
              &L13  & 11.8    & 0.8   &        "             & AT 6        \\
"   &             &                       &       &              &   &
              &L77  & 13.32   & 0.60  &        "             & E79           \\
"   &             &                       &       &              &   &
              &AT 50 & 15.21& 1.32  &        "             &               \\
"   &             &                       &       &              &   &
              &AT 60 & 15.52& 0.95  &        "             &               \\
"   &             &                       &       &              &   &
              &AT 119& 16.93& 0.86  &        "             &               \\
12  & 17h24m39.1s &-49$^\circ$56\am 54\as & 12\as & 1.0$\pm$ 0.3 & H &
              &E51  & 14.82   & 0.74  & $1.6 \cdot 10^{30}$ & AT 45       \\
13  & 17h25m10.2s &-49$^\circ$55\am 59\as & 10\as & 1.5$\pm$ 0.4 & H &\\
14  & 17h25m 5.3s &-49$^\circ$56\am 44\as &  7\as & 3.1$\pm$ 0.5 & H &
              &L44  & 10.54   & 0.27  & $5.0 \cdot 10^{30}$ & BS (A5)       \\
15  & 17h23m37.6s &-49$^\circ$57\am 16\as & 13\as & 1.4$\pm$ 0.4 & H &
               &OUT  &$\sim$14 &       &         & POSS     \\
16  & 17h26m 9.8s &-49$^\circ$58\am  3\as & 14\as & 2.0$\pm$ 0.5 & H &
               &OUT  &$\sim$16 &       &         & POSS     \\
17  & 17h23m25.6s &-49$^\circ$58\am  9\as & 13\as & 1.9$\pm$ 0.4 & H &\\
18  & 17h24m36.6s &-49$^\circ$59\am 17\as & 10\as & 1.6$\pm$ 0.4 & H &
              &AT 4218&15.15& 1.06  & $2.6 \cdot 10^{30}$ &               \\
"   &             &                       &       &              &   &
              &AT 4219&15.81& 0.76  &        "             &               \\
19  & 17h25m55.3s &-50$^\circ$ 0\am  5\as &  6\as & 7.5$\pm$ 0.8 & H &
0.19$\pm$0.11\\
20  & 17h26m20.2s &-50$^\circ$ 1\am 24\as & 14\as & 2.4$\pm$ 0.5 & H &\\
21  & 17h24m42.1s &-50$^\circ$ 1\am 42\as & 10\as & 1.3$\pm$ 0.6 & H &\\
22  & 17h23m46.2s &-50$^\circ$ 2\am 52\as &  7\as & 5.0$\pm$ 0.6 & H &
               &OUT  & 11.65   &       &         & GSC      \\
23  & 17h23m37.7s &-50$^\circ$ 3\am 15\as &  6\as &11.3$\pm$ 0.9 & H &
0.49$\pm$0.08\\
24  & 17h25m45.2s &-50$^\circ$ 4\am 20\as & 17\as & 4.3$\pm$ 0.7 & H &
               &OUT  & $>$16   &       &         & POSS     \\
25  & 17h25m 0.1s &-50$^\circ$ 5\am 11\as &  5\as & 6.7$\pm$ 0.7 & H &
0.33$\pm$0.11  &OUT  &$\sim$13.5&      &         & POSS     \\
\end{tabular}
\end{flushleft}   
\end{table*}
 
\section{Results}

\subsection{Identifications}

As mentioned in the previous section, eight X-ray sources could be identified
with GSC stars.
In order to obtain identifications with possible members of the cluster,
we cross-correlated our source list with the optical catalogs of stars
in the field of IC\,4651. Four such lists have been presented:
Eggen (1971), Lindoff (1972), Anthony-Twarog \& Twarog (1987), and
Anthony-Twarog et al. (1988). We consider a tentative identification if
a catalogued star falls within the 90\% error box of an X-ray source
and no other stars are present in the error box. From these four lists
we found an unambiguous possible counterpart to four X-ray sources.
Two additional sources (\#11 and \#18) have more than one optical star
within their error boxes, complicating the identification procedure.
For the sources that cannot be identified in this way and for those
falling outside the regions covered in the references mentioned above,
we used the Guide Star Catalog and examined the Digitized Sky Survey
(Postman et al., in preparation). The proposed identifications 
are summarized in Table 1.

Currently there are neither
radial velocity nor proper motion studies published in the literature
for this cluster, and therefore no firm determination of membership is
available. However, the relative compactness and richness of the cluster 
(a hundred stars within an angular
diameter of $\sim 15 '$) ensures that most of the stars within this radius
are indeed members, as can be easily seen from the color-magnitude
diagram (Fig. 1). Given the angular size of the cluster, it is unlikely
that those X-ray sources outside the region covered by optical studies
are related to the cluster, even if they have a relatively bright
optical counterpart.
Anthony-Twarog \& Twarog (1987) derive an indication of membership
from ubvyH$_\beta$ photometry. In this way, they can separate probable
single members from non-members and binaries, but cannot discriminate
between the latter two.

As can be seen from Table 1, six X-ray sources have possible counterparts
listed in Lindoff (1972), Eggen (1971) or Anthony-Twarog et al. (1988).
Source \#8 is identified with the red giant L241; this star is
included in a list of five binaries discovered by Mermilliod et al.
(1995) in a sample of 20 red giants in the IC\,4651 field.
It is a single line spectroscopic binary with a period of 75.17 days
and an eccentricity of $0.09 \pm 0.02$. Source \#10 is identified with
L234 and lies on the cluster main sequence (see Fig. 1), as does
source \#12, identified with E51. Source \#14, is identified with the blue
straggler L44. Anthony-Twarog \& Twarog (1987) find it to be a binary
member of the cluster on the basis of their ubvyH$\beta$ photometry.
In the X-ray error box of source \#18 there are two weak stars studied
by Anthony-Twarog et al. (1988). In Fig. 1 the two possible counterparts 
are represented by triangle symbols. Finally, the position of source \#11
coincides with a conglomerate of at least six stars. In Fig. 1 we
plot the three brightest of these, which we regard as the most likely
candidates candidates, as crosses.
All the sources marked in 
Table 1 with `OUT' as optical ID are outside the regions studied by the 
above authors and probably
are not cluster members. We identified four of them with stars
from the Guide Star Catalog, one with a SAO star, while to the other six
we tentatively assigned a magnitude using the Digitized Sky Survey 
(Postman et al., in preparation).
The remaining eight X-ray sources do not have visible optical candidates in
their error boxes and are not listed in Table 1; two of them (\#13,21)
lie in the central part of the cluster.

\begin{figure*}
\includegraphics{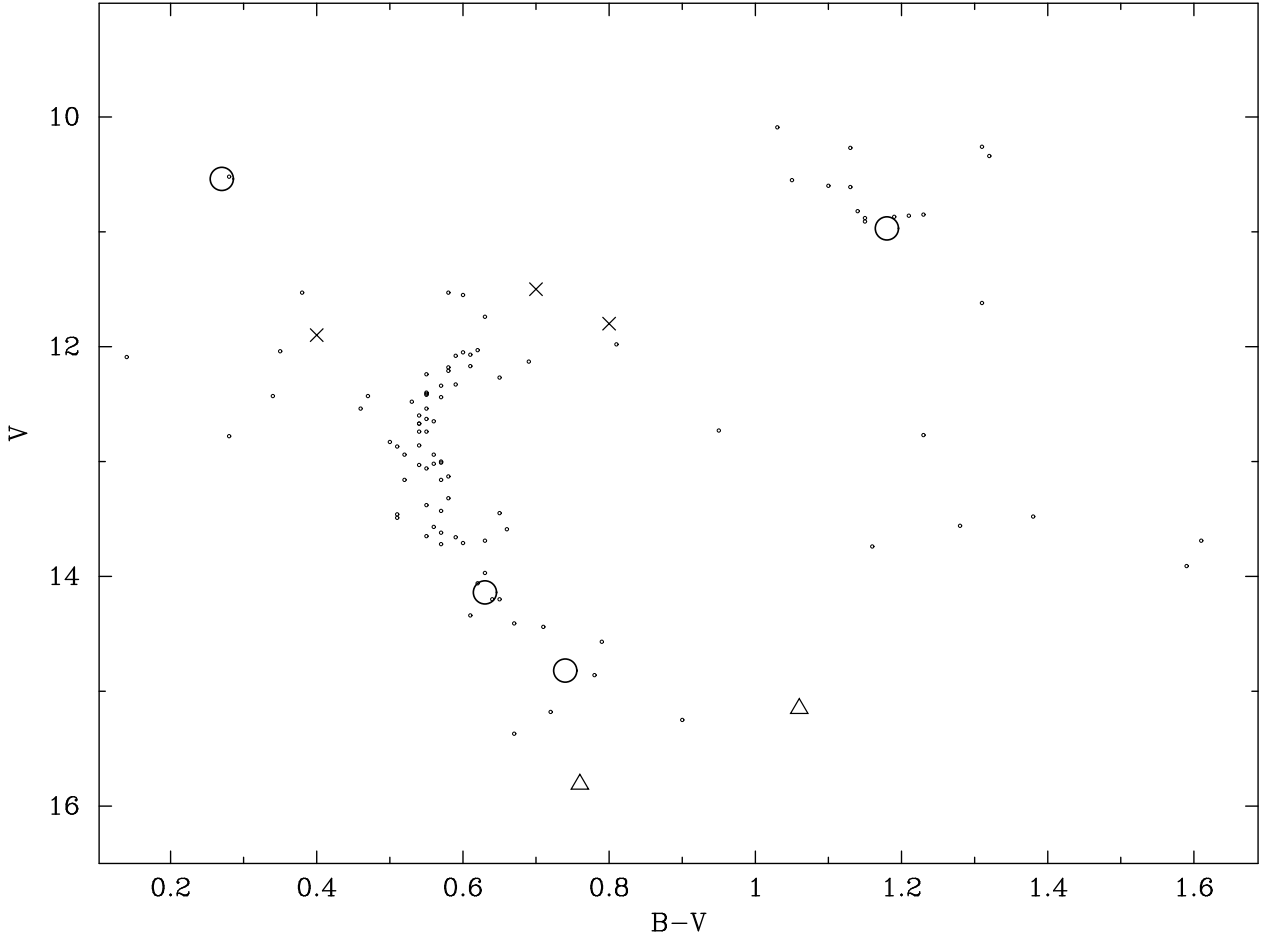}\\
\caption{
Color-magnitude diagram for IC\,4651 (data from Anthony-Twarog et al.
1988). Circles are X-ray detections with a likely member,
crosses and triangles are stars within the X-ray error box of sources
\#11 and \#18, respectively.}

\end{figure*}

IC\,4651 is located at low galactic latitude (b$_{II} \sim -8^\circ$). The
total galactic interstellar absorption along the line of sight, estimated 
from radio data, is
N$_H \sim 1.7\times 10^{21}$cm$^{-2}$ (Dickey \& Lockman 1990). 
To estimate the number of field sources expected, we assumed a
typical power-law spectrum with photon index 2.0 and the above
value of N$_H$. With these parameters our limiting count rate of $\sim$1
ct/ksec translates into a flux limit of 2.0$\times 10^{-14}$\,erg
cm$^{-2}$s$^{-1}$ (in the 0.4--2.5 keV band). From the log N--log S
distribution derived from the ROSAT Lockman hole deep survey 
(Hasinger et al. 1997), we estimate
that roughly 15 detections of extragalactic sources are
expected in our observation. All the sources without a POSS optical 
counterpart are therefore most likely extragalactic, as well as some of 
the weak identifications outside the region of the cluster.

\subsection{X-ray properties}

The number of counts detected from most of the sources is too low to
produce meaningful hardness ratios. 
All sources but three are not detected in the soft
ROSAT band, due to the relatively large value of the interstellar
absorption in front of the cluster.
In order to estimate the X-ray luminosity in the ROSAT band for the
detected members, we adopted a spectral model typical of RS CVn binaries.
Following Dempsey et al. (1993a), who studied the full sample of RS CVn
binaries detected in the RASS, we used a two-temperature thin emission
plasma model (according to Raymond \& Smith 1977). The parameters used
are the average of the values in Dempsey et al. (1993a): ${\rm kT}_{low}$=
0.175 keV, ${\rm kT}_{high}$=1.4 keV, ${\rm EM}_{high}/{\rm EM}_{low}$=6. 
For each star we used a value of interstellar absorption 
N$_H$ = $5.6 \times 10^{20}$, as derived from the most recent estimate
of E(B--V)=0.09 (see Anthony-Twarog \& Twarog 1987). The derived X-ray 
luminosities, assuming a distance of 850\,pc (Eggen 1971, Anthony-Twarog
\& Twarog 1987), are reported
in Table 1. Although the detections correspond to the hard PSPC band
(0.4--2.4 keV), the values are given for the full PSPC range 0.1--2.4
keV to allow comparison with other systems. We estimate that the
uncertainties on the derived luminosities, due to a different value of
the temperatures and/or EM ratio but still within the typical range
observed for coronal sources, are up to 50\%.

For the five brightest X-ray sources we calculated the value of the
ROSAT hardness ratio HR2=(D--C)/(D+C), where C and D are the counts
in the channel range 52--90 (0.5--0.9 keV) and 91--201 (1.0--2.4 keV)
respectively (see Table 1). 
We expect a value of 0.32 from the model adopted for the
flux conversion (see above).
As one can see, the detected binary (\#8) has a HR2 value compatible
with the expectation from active binary sources. 

\section{Discussion}

The age of IC\,4651 is about twice that of NGC\,6940 and similar to that 
of NGC\,752. As in the case of NGC\,752 (Belloni \& Verbunt 1996), we expect
to detect coronal sources only if they are in binary systems; moreover,
due to saturation effects one would expect to detect more easily giant
stars than main sequence stars (see discussions in Stauffer et al. 1994;
Belloni \& Verbunt 1996; Belloni \& Tagliaferri 1997).
As can be seen from Fig. 1, in IC\,4651 there are more than a dozen giants.
Five of them are found to be binaries by Mermilliod et al. (1995).
However, only one of them has a period short enough to sustain a high
level of stellar activity, and this is the binary we detect as an
active X-ray source. 
In this scenario, one expects the stars to co-rotate and therefore
the orbit to be circular. This binary is slightly eccentric (e=0.1),
but its characteristics are compatible with
co-rotation (see Verbunt \& Phinney 1995). This is also the brightest 
source of the six that we suppose to be at the cluster distance, in 
agreement with the presence of a saturation level, which is higher
for evolved stars (see Belloni \& Verbunt 1996).

In NGC\,6940 we had not detected main sequence stars,
but only giants. Of these, two are not classified as binaries; one
in particular has been extensively studied (Mermilliod \& Mayor 1989)
and is very unlikely to be a binary (see Belloni \& Tagliaferri 1997).
Thus, their detection in the X-ray band is puzzling.
On the contrary, in IC~4651 of all known giants only one, a binary, is 
detected. Two other stars lying on the cluster main sequence are also
detected. Given the age of the cluster, we expect them to be binaries.
At the bottom end of the cluster main sequence, somewhat offset from it,
lie the two possible counterparts of source \#18. They could be either
field stars or weak members of the cluster.

In the error box of source \#11, which is only 13 arcsec, there are at least
seven stars, which look like a sub-cluster.
The three brightest of these stars are plotted as crosses in Fig. 1.
If these stars are all members of the cluster, then one is a blue straggler
(i.e. it lies in a region of the color-magnitude diagram occupied by blue 
stragglers), while the other two are stars in the process
of evolving toward the red giant branch. The possibility cannot be ruled out
that more than one of these stars contribute to the
detected X-ray emission.

Finally, source \#14 is identified with a blue straggler, found
by Anthony-Twarog \& Twarog (1987) to be a binary. This is the third blue
straggler detected in the X-ray band, the other two being in M67 and
NGC\,752 respectively. Of all blue stragglers known in the old and
intermediate-age open clusters that we studied, only three
are detected in X rays. All three have
measurements that indicate binarity (see Belloni 1998),
and once again this is probably the key to X-ray emission. These are not
the only binaries know among the blue stragglers in these clusters, 
showing that
binarity does not imply strong stellar activity per se.
One of the two stars also has to be of late spectral type in order
to have an enhanced dynamo activity, like in Algol systems.

For the sources marked with `OUT'  in Table 1 we have no color information;
however, from their magnitudes, most of them should be stars.
They lie outside the inner 10 arcmin diameter region of the cluster
and are probably non-members. We have already planned follow-up optical
observations to determine the physical nature of all detected X-ray sources
in the IC\,4651 field.

\acknowledgements{GT acknowledge partial support from the Italian Space
Agency. We thank F. Verbunt for useful discussions.}

\section{References}
\def\xr#1{\parindent=0.0cm\hangindent=1cm\hangafter=1\indent#1\par}     
\def\aa#1#2{{Acta Astron.} { #1}, {#2}.}
\def\aaa#1#2{{A\&A} { #1}, {#2}.}
\def\aas#1#2{{A\&AS} { #1}, {#2}.}
\def\aar#1#2{{A\&AR} { #1}, {#2}.}
\def\aj#1#2{{AJ} { #1}, {#2}.}
\def\apj#1#2{{ApJ} { #1}, {#2}.}
\def\apl#1#2{{ApJ} { #1}, L{#2}.}
\def\aps#1#2{{ApJS} { #1}, {#2}.}
\def\mn#1#2{{MNRAS} { #1}, {#2}.}
\def\nat#1#2{{Nat} { #1}, {#2}.}
\def\pasj#1#2{{PASJ} { #1}, {#2}.}
\def\pasp#1#2{{PASP} { #1}, {#2}.}
\def\ssr#1#2{{Space Sci. Rev.} { #1}, {#2}.}
\par
\xr{Anthony-Twarog B.J., Twarog B.A., 1987, AJ 94, 1222}
\xr{Anthony-Twarog B.J., Mukherjee K., Caldwell N., Twarog B.A., 1988,
AJ 95, 1453}
\xr{Belloni T. 1998, in ``Cool Stars in Cluster and Associations:
magnetic activity and age indicators", Mem. SAIt in press}
\xr{Belloni T., Tagliaferri G., 1997, A\&A ,326, 608}
\xr{Belloni T., Verbunt F., 1996, \aaa {305}{806}}
\xr{Belloni T., Verbunt F., Schmitt J.H.M.M., 1993,
    \aaa {269}{175}}
\xr{Caillault J.-P., Helfand D.J., 1985, \apj {289}{279}}
\xr{Cruddace R.G., Hasinger G.R., Schmitt J.H.M.M., 1988,
    in {``Astronomy from large databases''}, eds. Murtagh F. and
    Heck A., 177}
\xr{Dempsey R.C., Linsky J.L., Schmitt J.H.M.M., Fleming T.A., 1993a,
    \apj {413}{333}}
\xr{Dempsey R.C., Linsky J.L., Fleming T.A., Schmitt J.H.M.M., 1993b,
    \aps {86}{599}}
\xr{Dickey J. M., Lockman F. J., 1990, {ARA\&A} {28}, 215}
\xr{Eggen O.J, 1971, ApJ 166, 87}
\xr{Hasinger G., Burg R., Giacconi R., Schmidt M., Tr\"umper J.,
Zamorani G., 1997, A\&A in press}
\xr{Lasker B.M., Sturch C.R., McLean B.J., et al., 1990, AJ 99, 2019}
\xr{Lindoff U., 1972, \aas {7}{231} }
\xr{Mermilliod J.-C., Andersen J., Nordstr\"om B., Mayor M., 1995, 
    \aaa {299}{53} }
\xr{Mermilliod J.-C., Mayor M., 1989, \aaa{219}{125}}
\xr{Micela G., Sciortino S., Vaiana G.S., Schmitt J.H.M.M., Stern R.A.,
	Harnden F.R., Rosner R., 1988, \apj {325}{798}}
\xr{Micela G., Sciortino S., Vaiana G.S., Rosner R.,
    Schmitt J.H.M.M., 1990, ApJ 348, 557}
\xr{Pallavicini R., Golub L., Rosner R., et al., 1981, \apj {248}{279} }
\xr{Pasquini L., Belloni T., 1994, in ``Cool Stars, Stellar Systems,
    and the Sun'', J.-P. Caillault (ed.), ASP Conference Series, {64}, 122}
\xr{Patten B.M., Simon T., 1993, ApJ 415, L23}
\xr{Randich S. 1997, in ``Cool Stars in Cluster and Associations:
magnetic activity and age indicators", Mem. SAIt in press}
\xr{Randich S., Schmitt J.H.M.M., 1995, A\&A 298, 115}
\xr{Randich S., Schmitt J.H.M.M., Prosser C.F., Stauffer J.R.,
    1995, A\&A 300, 134}
\xr{Randich S., Schmitt J.H.M.M., Prosser C.F., Stauffer J.R.,
    1996a, A\&A 305, 785}
\xr{Randich S., Schmitt J.H.M.M., Prosser C.F., 1996b, A\&A 313, 815}
\xr{Pfeffermann E., Briel U.G., Hippmann H., et al., 1986, {SPIE}, {733}, 519}
\xr{Raymond J., Smith B., 1977, \aps {35}{419}}
\xr{Schmitt J.H.M.M., Micela G., Sciortino S., Vaiana G.S., Harnden F.R.,
	Rosner R., 1990, \apj {351}{492}}
\xr{Skumanich A., 1972, \apj {171}{565}}
\xr{Stern R.A., Zolcinsky M.C., Antiochos S.C., Underwood J.M., 1981,
    \apj {249}{647} }
\xr{Stern R.A., Schmitt J.H.M.M., Pye J.P., Hodgkin S.T., Stauffer J.,
    1992, ApJ 399, L159}
\xr{Stauffer J.R., Caillault J.-P., Gagn\'e M., Prosser C.F., Hartmann
    L.W., 1994, ApJS 91, 625}
\xr{Tr\"umper J., 1983, {\it Adv. Space Res.} {\bf 2}, no.4, 241.}
\xr{Verbunt F., Phinney, E.S., 1995, \aaa{296}{709}}
\xr{Zimmermann H.U., Becker W., Belloni T., et al., 1994, MPE Report 257.}

\end{document}